\documentclass[aps,pra,twocolumn,showpacs,floatfix,superscriptaddress]{revtex4-1}

  \usepackage[utf8]{inputenc}
  \usepackage[T1]{fontenc}
  \usepackage{float} 
  \usepackage{color}  

\usepackage{graphicx}
\usepackage{amsmath}
\usepackage{amssymb}
\usepackage{bbm}
\usepackage{indentfirst}
\usepackage{dcolumn}
\usepackage{soul}
\usepackage{mathrsfs}
\usepackage{tikz}
\usetikzlibrary{calc}
\usetikzlibrary{arrows.meta}

\begin{document}

\title{Temperature scaling of the Dzyaloshinsky--Moriya interaction in the spin wave spectrum}

\author{Levente R\'{o}zsa}
\email{rlevente@physnet.uni-hamburg.de}
\affiliation{Department of Physics, University of Hamburg, D-20355 Hamburg, Germany}
\affiliation{Institute for Solid State Physics and Optics, Wigner Research Centre for Physics, Hungarian Academy of Sciences,
P.O. Box 49, H-1525 Budapest, Hungary}
\author{Unai Atxitia}
\affiliation{Department of Physics, University of Konstanz, D-78457 Konstanz, Germany}
%\author{Rocio Yanes}
%\affiliation{University of Salamanca, E-37008 Salamanca, Spain}
\author{Ulrich Nowak}
\affiliation{Department of Physics, University of Konstanz, D-78457 Konstanz, Germany}
\date{\today}

\begin{abstract}
The temperature scaling of the micromagnetic Dzyaloshinsky--Moriya exchange interaction 
%spin stiffness
is calculated for the whole range of temperature. We use Green's function theory to derive the finite-temperature spin wave spectrum of ferromagnetic systems described by a classical atomistic spin model Hamiltonian. Within this model, we find universal expressions for the temperature scaling not only of the Dzyaloshinsky--Moriya interaction but also of the Heisenberg exchange stiffness and the single-ion anisotropy. In the spirit of multiscale models, we establish a clear connection between the atomistic interactions and the temperature-dependent coefficients in the spin wave spectrum and in the micromagnetic free energy functional. We demonstrate that the corrections to mean-field theory or the random phase approximation for the temperature scaling of Dzyaloshinsky--Moriya and Heisenberg exchange interactions assume very similar forms. In the presence of thermal fluctuations and Dzyaloshinsky--Moriya interaction an anisotropy-like term emerges in the spin wave spectrum which, at low temperature, increases with temperature, in contrast to the decreasing single-ion anisotropy. We evaluate the accuracy of the theoretical method by comparing it to the spin wave spectrum calculated from Monte Carlo simulations. 
\end{abstract}

\maketitle

\section{Introduction}

Chirality in magnetic systems appears due to the broken inversion symmetry of the crystal. Microscopically it stems from a relativistic exchange interaction between magnetic moments known as the Dzyaloshinsky--Moriya interaction\cite{Dzyaloshinsky,Moriya}. Originally introduced to account for the weak magnetic moment of some antiferromagnetic systems\cite{Dzyaloshinsky}, it has been demonstrated that this type of coupling gives rise to chiral spin structures ranging from domain walls\cite{Thiaville,Ryu} through spin spirals\cite{Bode,Meckler} to magnetic skyrmions\cite{Bogdanov,Muhlbauer,Yu2}. As Dzyaloshinsky--Moriya interaction results from spin-orbit interactions, it represents a substantial aspect of a new field of research called spin-orbitronics, with potential applications in future spintronic devices\cite{Allwood,Parkin,Fert}.

Besides influencing static spin configurations, the Dzyaloshinsky--Moriya interaction also lends a chiral character to the magnetic excitations of the system, which are known as magnons or spin waves. In ferromagnetic systems, this is observable in the shift of the minimum of the parabolic spin wave dispersion relation away from the $\boldsymbol{k}=\boldsymbol{0}$ point, thereby lifting the energy degeneracy between magnons propagating in opposite directions\cite{Melcher,Udvardi}. Common experimental methods for determining this asymmetry include neutron scattering for bulk magnets\cite{Coldea,Sato}, and Brillouin light scattering\cite{Nembach,Belmeguenai}, spin-polarized electron energy loss\cite{Zakeri,Zakeri3,Zakeri2} and propagating spin wave spectroscopy\cite{Lee} for thin films. Recent theoretical investigations based on the micromagnetic model have proposed magnonic devices based on the chiral character of spin waves\cite{Garcia-Sanchez,WYu}.

In spin glasses it has been demonstrated that the addition of nonmagnetic heavy metal impurities enhances the anisotropy field. This effect has also been attributed to the presence of the Dzyaloshinsky--Moriya interaction in connection with the noncollinear alignment of the spins in spin glasses\cite{Fert2,Levy}.
The chiral interaction is also responsible for the canting of spins at the edges of nanomagnets, which also induces an anisotropy field competing with demagnetization effects\cite{Cubukcu}. However, the Dzyaloshinsky--Moriya interaction does not influence the orientation of the ground state in extended ferromagnetic systems\cite{Melcher,Udvardi}, since in this case all spins are parallel to each other.

Significant research attention has been turned towards investigating phase transitions\cite{Muhlbauer,Yu2} and determining the lifetime of metastable spin structures\cite{Hagemeister,Oike,Rozsa2} in chiral systems at finite temperature. The microscopic background of such relaxation mechanisms is the thermal fluctuation of localized magnetic moments, leading to a reduced magnetization at higher temperature. Micromagnetic models rely on the approximation that the magnetization is only slowly varying over the sample, leading to an effective averaging of the magnetic moments over small volumes. During this averaging, it is necessary to take into account the temperature dependence of the magnetization as well as that of the effective interaction parameters. These effective interaction parameters are crucial for understanding phase transitions and lifetimes of metastable states.

Notably, dynamic properties can be calculated through finite-temperature approaches such as the Landau--Lifshitz--Bloch equation\cite{Chubykalo-Fesenko,Atxitia2} that fundamentally rely on temperature-dependent micromagnetic parameters. Such approaches are especially important in the emerging field of spin caloritronics, concerning the interplay between spin, charge and heat degrees of freedom. It has been established that domain wall motion in ferromagnets\cite{Schlickeiser} and antiferromagnets\cite{Selzer} under thermal gradients is dominated by the so-called entropic torque, defined by the temperature derivative of the Heisenberg exchange stiffness. It is expected that further dynamical effects appear in the presence of the Dzyaloshinsky--Moriya interaction. For instance, in the field of ultrafast spin dynamics 
%, where fs laser pulse strongly heats the magnetic system in the ps timescale, 
the emergence of metastable magnetic textures was demonstrated, such as vortex-antivortex pairs in Fe thin films\cite{Eggebrecht} or skyrmions in thin TbFeCo films\cite{Finazzi}. Therefore, it is important to develop theoretical methods for the calculation of effective temperature-dependent parameters.

Analytical results for specific types of interaction parameters are available in the literature, usually given as a power law of the magnetization $m^{\kappa}$. The most well-known example is the $\kappa=l(l+1)/2$ power law for $l$th order magnetic anisotropy\cite{Callen2}, although deviations from this behavior can also be found in certain systems\cite{Okamoto,Asselin}. For the Heisenberg exchange interaction, most applications use the result of mean-field theory\cite{Nembach}, where it scales with the second power of the magnetization $m^{2}$; however, it has been demonstrated that corrections to this approximation are necessary in most systems\cite{Heider,Atxitia}.

In comparison, the temperature dependence of the Dzyaloshinsky--Moriya interaction seems to be less explored, and the implicit estimations provided so far do not fully agree with each other. For instance, in Ref.~\cite{Barker} it was concluded that the temperature dependence of the size of antiferromagnetic skyrmions may be described by an expression containing temperature-independent Dzyaloshinsky--Moriya interactions. In contrast, a significant softening of the chiral interaction was reported for ultrathin ferromagnetic films in Ref.~\cite{Kim}. In Ref.~\cite{Hasselberg}, it was found that the wavelength of spin spirals, proportional to the ratio between the Heisenberg and Dzyaloshinsky--Moriya interactions, is independent of the temperature, implying a similar temperature scaling for the two terms. The independence of the period of noncollinear order on the temperature has also been demonstrated experimentally in several systems\cite{Sonntag,Sessi,Bergmann,Shibata}. Solving this kind of apparent discrepancy 
% also of interest for its implications in spin caloritronics in non-trivial topological spin textures. This undertaking 
requires theoretical methods that directly provide the temperature scaling of both Heisenberg and Dzyaloshinsky--Moriya micromagnetic exchange interactions. 

Preliminary results along this line were provided in Ref.~\cite{Rozsa}, where the spin wave spectrum was calculated for a ferromagnetic monolayer containing Dzyaloshinsky--Moriya interaction, and the softening of the frequencies at low temperature was described by linear spin wave theory. Green's function theory in statistical mechanics\cite{Callen} provides a more robust theoretical framework for the description of thermal spin fluctuations and the finite-temperature spin wave spectrum over wide temperature ranges in both quantum and classical systems\cite{Bastardis,Campana}. Originally developed for the determination of the temperature scaling of the magnetization, the method was naturally extended for calculating effective temperature-dependent interaction parameters\cite{Nakamura,Bastardis}. The application of Green's function theory to the Dzyaloshinsky--Moriya interaction\cite{You,Cho} so far has been restricted to the random phase approximation\cite{Tyablikov}, which neglects the corrections appearing due to correlations between the spins\cite{Atxitia,Bastardis}.

In this paper, we investigate the temperature dependence of the Dzyaloshinsky--Moriya interaction in a ferromagnet. Using Callen's formulation of Green's function theory\cite{Callen}, we find significant corrections to mean-field theory or  random phase approximation due to transversal spin fluctuations. These corrections assume very similar form for the Heisenberg and Dzyaloshinsky--Moriya interactions, in agreement with the microscopic description which derives the two quantities from the same principle\cite{Moriya,Fert2,Nembach}. Furthermore, we demonstrate that the Dzyaloshinsky--Moriya interaction induces an anisotropy-like term which increases the spin wave frequency at zero wave vector, an effect which is only observable at finite temperature in ferromagnets. By comparing the results to Monte Carlo simulations, we demonstrate that the theory successfully accounts for most of the fluctuation corrections.
%, while for the anisotropy the behavior is significantly different. 
%{\del{We explicitly calculate the spin wave spectrum for an ultrathin film, which kind of system plays an important role in suggested applications based on the Dzyaloshinsky--Moriya interaction}}\cite{Fert,Garcia-Sanchez}{\del{. Furthermore, it is expected that the transversal spin fluctuations responsible for the corrections beyond mean-field theory play a pronounced role in two-dimensional systems; it is known that these transversal fluctuations are responsible for the disappearance of long-range order at finite temperature in isotropic systems}}\cite{Mermin}{\del{. By comparing the results to Monte Carlo simulations, we demonstrate that the theory accounts for most of the fluctuation corrections.}}

\section{Finite-temperature spin wave spectrum\label{sec2}}

\subsection{Green's function theory}

For the description of the magnetic system, we introduce the classical atomistic spin Hamiltonian
\begin{eqnarray}
H=&&-\frac{1}{2}\sum_{i,j}J_{ij}\boldsymbol{S}_{i}\boldsymbol{S}_{j}-\frac{1}{2}\sum_{i,j}\boldsymbol{D}_{ij}\left(\boldsymbol{S}_{i}\times\boldsymbol{S}_{j}\right)\nonumber
\\
&&-\sum_{i}K^{zz}\left(S_{i}^{z}\right)^{2}-\mu_{s}\sum_{i}B^{z}S_{i}^{z}.\label{eqn1}
\end{eqnarray}

Here the $\boldsymbol{S}_{i}$ variables denote unit vectors, $J_{ij}$ is the Heisenberg exchange interaction between atoms at sites $i$ and $j$, $\boldsymbol{D}_{ij}$ is the Dzyaloshinsky--Moriya vector, $K^{zz}$ is the single-ion magnetocrystalline anisotropy, $\mu_{s}$ is the magnetic moment, and $B^{z}$ is the external magnetic field. The number of spins in the lattice will be denoted by $N$. We will assume that the ground state of the system is ferromagnetic along the $z$ direction. The interaction coefficients in Eq.~(\ref{eqn1}) are determined by microscopic electronic processes such as the overlap between wave functions (direct exchange) or hopping processes (superexchange\cite{Anderson,Moriya}). In a multiscale description\cite{Atxitia}, the coefficients in Eq.~(\ref{eqn1}) may be determined from \textit{ab initio} calculations. It is not possible to consider the interplay between the noncollinear spin arrangement and the electronic structure\cite{Szilva} in the simple model presented here; therefore, we will suppose that $J_{ij},\boldsymbol{D}_{ij},K^{zz}$ are independent of temperature on the scale where magnetic ordering occurs.

%See footnote in Ref\footnote{I think the order should be, first to present the SW spectrum expression, which does not depend on the form one calculates the magnetisation. Moreover the obtained expression for the frequencies is also valid for the quantum derivation (with some renormalisation of the coefficients), validity that could be outlined in the paper to later on justify the use of the classical version in order to compare to atomistic spin dynamics MC simulations. But we could let open the possibility of the quantum case. Then one needs to resort some equation to calculate the magnetisation, but it does not really depends on the SW spectrum, and different approaches have been discussed in the literature. It seems that for classical systems, this Langevin form works fine, and for quantum systems the Brillouin form does.}

For calculating the spectrum of spin wave excitations at finite temperature, we will use the classical Green's function formalism\cite{Callen,Bastardis}, which results in a set of self-consistency equations. For their derivation see Appendix~\ref{secS1}. The spin wave spectrum in Fourier space $\boldsymbol{k}$ reads
\begin{eqnarray}
\omega_{\boldsymbol{k}}\left(T\right)=\frac{\gamma}{\mu_{s}}\left(\mathcal{J}_{\boldsymbol{0}}-\mathcal{J}_{\boldsymbol{k}}-\textrm{i}\mathcal{D}_{\boldsymbol{k}}+2\mathcal{K}^{zz}+\mu_{s}B^{z}\right),\label{eqn9}
\end{eqnarray}
with $\gamma=\frac{ge}{2m_{e}}$ the electron's gyromagnetic ratio, and effective temperature-dependent interaction parameters $\mathcal{J}_{\boldsymbol{k}},\mathcal{D}_{\boldsymbol{k}},\mathcal{K}^{zz}$. The correspondence between these parameters and the interaction coefficients in the atomistic Hamiltonian Eq.~(\ref{eqn1}) is given by
\begin{align}
\mathcal{J}_{ij}=&mJ_{ij}+\frac{m}{2}J_{ij}\textrm{Re}\left<S_{j}^{+}S_{i}^{-}\right>,\label{eqn10}
\\
\mathcal{D}_{ij}=&mD_{ij}+\frac{m}{2} J_{ij}\textrm{Im}\left<S_{j}^{+}S_{i}^{-}\right>,\label{eqn11}
\\
\mathcal{K}^{zz}=&K^{zz}\left(m-\frac{m}{2}\left<S_{i}^{+}S_{i}^{-}\right>\right)+\frac{m}{4}\sum_{j}D_{ij}\textrm{Im}\left<S_{j}^{+}S_{i}^{-}\right>,\label{eqn12}
\end{align}
where $D_{ij}$ denotes the $z$ component of the Dzyaloshinsky--Moriya vectors $\boldsymbol{D}_{ij}$.

Equations (\ref{eqn9})-(\ref{eqn12}) must be solved self-consistently together with the temperature scaling of the magnetization
\begin{eqnarray}
m=\left<S^{z}\right>=\coth\frac{1}{\Phi}-\Phi,\label{eqn13}
\end{eqnarray}
where
\begin{eqnarray}
\Phi=\frac{1}{N}\sum_{\boldsymbol{k}}\frac{\gamma}{\mu_s}\frac{k_{\textrm{B}}T}{\omega_{\boldsymbol{k}}}.\label{eqn14}
\end{eqnarray}

The transversal correlation function is given by
\begin{eqnarray}
\left<S^{-}_{\boldsymbol{k}}S^{+}_{-\boldsymbol{k}}\right>=\frac{2m}{N}\frac{\gamma}{\mu_s}\frac{k_{\textrm{B}}T}{\omega_{\boldsymbol{k}}}.\label{eqn15}
\end{eqnarray}

Following the multiscale description, Eqs.~(\ref{eqn10})-(\ref{eqn12}) will be used to determine the temperature-dependent interaction parameters in a micromagnetic model, which is based on a continuum free energy functional. For simplicity, here we will only consider spin modulations along the $x$ direction which is perpendicular to the magnetization; a generalization to more spatial dimensions can be found in Appendix~\ref{secS2}. The free energy density is given by
\begin{align}
f=\sum_{\alpha}\mathscr{A}\left(\partial_{x}S^{\alpha}\right)^{2}+\mathscr{D}L\left(\boldsymbol{S}\right)
-\mathscr{K}^{zz}\left(S^{z}\right)^{2}-\mathscr{M}B^{z}S^{z},\label{eqn16}
\end{align}
%appropriate in the long-wavelength approximation.
where $\boldsymbol{S}$ denotes the unit length spin vector field,
\begin{eqnarray}
L\left(\boldsymbol{S}\right)=S^{z}\partial_{x}S^{x}-S^{x}\partial_{x}S^{z}\label{eqn16a}
\end{eqnarray}
is the linear Lifshitz invariant\cite{Dzyaloshinsky}, and $\mathscr{A}, \mathscr{D}, \mathscr{K}^{zz}$ are the micromagnetic effective Heisenberg exchange, Dzyaloshinsky--Moriya interaction, and anisotropy, respectively. The magnetization density reads
\begin{eqnarray}
\mathscr{M}(T)=\frac{\mu_{s}}{\upsilon_{\textrm{\tiny WS}}} m(T),\label{eqn17}
\end{eqnarray}
where $\upsilon_{\textrm{\tiny WS}}$ is the Wigner--Seitz volume occupied by a single atom in the lattice.
%For simplicity we only considered modulations of the spins along the $x$ direction in Eq.~(\ref{eqn16}); in this case, the coefficients may be calculated from the atomistic parameters when the spin wave spectrum Eq.~(\ref{eqn9}) is calculated along the $x$ direction.

%The equation of motion for the continuum model can be constructed analogously to the lattice model\cite{Landau}, yielding
%\begin{eqnarray}
%\dot{\boldsymbol{S}}=\frac{\gamma}{\mathscr{M}}\boldsymbol{S}\times\frac{\delta F}{\delta \boldsymbol{S}}.\label{eqn18}
%\end{eqnarray}

%To illustrate the correspondence between Eq.~(\ref{eqn2}) and Eq.~(\ref{eqn18}), note that both reproduce the Larmor frequency $\gamma B^{z}$ in the noninteracting case.

Unlike the $J_{ij},D_{ij},K^{zz},$ and $\mu_{s}$ parameters in the atomistic model, the $\mathscr{A},\mathscr{D},\mathscr{K}^{zz},\mathscr{M}$ coefficients appearing in the micromagnetic model are temperature-dependent. Their importance lies in the fact that they are directly related to experimentally observable quantities such as the macroscopic magnetization, the domain wall width ($\delta\propto\sqrt{\mathscr{A}/\mathscr{K}^{zz}}$), skyrmion radius, or spin spiral wavelength. Furthermore, in the presence of temperature gradients it has been shown that analytical expressions for thermomagnonic torques can be directly derived from the temperature dependence of the micromagnetic Heisenberg exchange\cite{Schlickeiser,Selzer}.
%Yet the connection between atomistic and micromagnetic quantities is far from having a standard procedure. Implicit methods have been so far proposed, for instance, 

Several methods have been proposed for connecting the atomistic and micromagnetic parameters. For instance, $\mathscr{A}$ and $\mathscr{K}^{zz}$ can be calculated via the temperature dependence of the domain wall width and free energy in an implicit way\cite{Atxitia}. 
In this paper, we connect the quantities by comparing the spin wave spectrum obtained from the two approaches. 
%The effectively temperature-dependent interaction parameters calculated in Eqs.~(\ref{eqn10})-(\ref{eqn12}) may be used to determined the micromagnetic parameters in Eq.~(\ref{eqn16}) by the analogy between the spin wave frequencies.
For spin waves propagating along the $x$ direction, expanding Eq.~(\ref{eqn9}) for long wavelengths (small $\boldsymbol{k}$) yields the correspondence
\begin{eqnarray}
\mathscr{A}&=&\frac{1}{4}\frac{m}{\upsilon_{\textrm{\tiny WS}}}\sum_{\boldsymbol{R}_{i}-\boldsymbol{R}_{j}}\mathcal{J}_{ij}\left(x_{j}-x_{i}\right)^{2},\label{eqn19}
\\
\mathscr{D}&=&-\frac{m}{2\upsilon_{\textrm{\tiny WS}}}\sum_{\boldsymbol{R}_{i}-\boldsymbol{R}_{j}}\mathcal{D}_{ij}\left(x_{j}-x_{i}\right),\label{eqn20}
\\
\mathscr{K}^{zz}&=&\frac{m}{\upsilon_{\textrm{\tiny WS}}}\mathcal{K}^{zz},\label{eqn21}
\end{eqnarray}
where $\boldsymbol{R}_{i}=(x_{i},y_{i},z_{i})$ stands for the position of the spin $i$ in the lattice. Thus, $\left(x_{j}-x_{i}\right)$ corresponds to the distance between the spins $i$ and $j$ along the $x$ axis.
%{\disp Comment Unai: here, at least in the first relation, the common relation has a $1/2$ instead of a $1/4$. see Atxitia et al. PRB 2010, definition of the micromagnetic exchange at the beginning of the paper and compare.}

\subsection{Discussion}

Equations (\ref{eqn10})-(\ref{eqn12}) together with (\ref{eqn19})-(\ref{eqn21}) constitute the main results of this paper. Using these expressions, it is possible to directly connect first principles calculations to micromagnetic models in a multiscale approach. This procedure may enable bypassing time-consuming atomistic spin model simulations for the determination of micromagnetic parameters. %; the accuracy of the analytical expressions is investigated in Sec.~\ref{sec3}
The calculations may also be generalized to quantum spins, 
%{\disp [ https://doi.org/10.1103/PhysRevB.83.165114]}
which modifies the expressions Eqs.~(\ref{eqn13})-(\ref{eqn15}) for the self-consistency by accounting for quantum statistics\cite{Callen} instead of the classical statistical limit considered here, but leaves Eqs.~(\ref{eqn10})-(\ref{eqn12}) for the effective parameters essentially unchanged.

The accuracy of Eqs.~(\ref{eqn10})-(\ref{eqn12}) is determined by the fluctuation corrections, which were first suggested to be included by Callen\cite{Callen}, with the appropriate classical limit given in Refs.~\cite{Bastardis,Campana}. This is encapsulated in the terms proportional to the transversal correlation function $\left<S_{j}^{+}S_{i}^{-}\right>$. Without this term, one would obtain $\mathscr{A},\mathscr{D},\mathscr{K}^{zz}\propto m^{2}$ in the micromagnetic description, corresponding to the random phase approximation\cite{Tyablikov} in the language of Green's functions. Neglecting the fluctuations may also be interpreted as a mean-field approximation\cite{Atxitia}.

%{\disp These equation together with Eqs 6 to 8 can directly be used to connect first principles calculations of the coefficients to micromagnetic theories bypassing computer consuming atomistic spin models.}
%{\disp Start discussion with some example: For instance, at zero temperature, for only $z$ n.n exchange interaction, one recovers the well known expression for the exchange stiffness $A=zJ/2a$. At finite temperature, the connection would be  $\mathscr{A}=A m^2(1+0.5\textrm{Re}\left<S_{j}^{+}S_{i}^{-}\right>)$, and in the MFA 
%$\mathscr{A}=A m^2$. This paragraph is hard to follow:}
%Equations~(\ref{eqn10})-(\ref{eqn12}) have an analogous form for the Heisenberg exchange, Dzyaloshinsky--Moriya interaction and single-ion anisotropy: a leading order term proportional to the magnetization $m$, and extra corrections due to the finite transversal correlation function $\left<S_{j}^{+}S_{i}^{-}\right>$. 
%The first term yields $\mathscr{A},\mathscr{D},\mathscr{K}^{zz}\propto m^{2}$ in the micromagnetic description; this can be understood as a mean-field theory\cite{Atxitia} or the random phase approximation\cite{Tyablikov} in the language of Green's function theory.

In the case of the single-ion anisotropy, it is long known that the correlation corrections play an important role; at low temperature, they modify the magnetization dependence of the micromagnetic anisotropy coefficient from $\mathscr{K}^{zz}\propto m^{2}$ to $\mathscr{K}^{zz}\propto m^{3}$\cite{Callen2}, a significantly faster decrease than in the random phase approximation. 

For the Heisenberg exchange interaction, the correlation correction has an opposite sign (cf. Eqs. (\ref{eqn10}) and (\ref{eqn12})), which leads to a slower decrease of the parameter $\mathscr{A}$ with temperature compared to the prediction of the random phase approximation. If the magnetization dependence of the exchange stiffness is expressed in the form of a power law $\mathscr{A}\propto m^{\kappa_{\mathscr{A}}}$ at low temperature, this yields $\kappa_{\mathscr{A}}<2$. As discussed in Ref.~\cite{Atxitia}, the exact value of the exponent depends on the system parameters, in particular the number of neighbors considered and the strength of the anisotropy. Note that Eqs.~(\ref{eqn10})-(\ref{eqn12}) describe the temperature dependence of the effective interaction coefficients for arbitrary pairs of atoms\cite{Swendsen}; since the correlation function $\left<S_{j}^{+}S_{i}^{-}\right>$ decays for further neighbors, the random phase approximation gives a better prediction in this case. Furthermore, the correlations decrease faster if the correlation length $\xi$ is smaller, which is directly connected to the spin wave frequency at zero wave vector by $\xi^{-2}\propto\omega_{\boldsymbol{0}}=\frac{\gamma}{\mu_{s}}\left(2\mathcal{K}^{zz}+\mu_{s}B^{z}\right)$. The correlation length is expected to play an especially important role in two-dimensional systems, where the fluctuations destroy long-range order at finite temperature in the absence of the spin wave gap\cite{Mermin}.

For the Dzyaloshinsky--Moriya interaction, Eqs.~(\ref{eqn10})-(\ref{eqn11}) demonstrate that the correlation correction has the same sign as in the case of the Heisenberg exchange interaction. Regarding the magnitude of the corrections in Eqs.~(\ref{eqn10})-(\ref{eqn11}), note that the correlation function $\textrm{Im}\left<S_{j}^{+}S_{i}^{-}\right>=\left<\left(\boldsymbol{S}_{j}\times\boldsymbol{S}_{i}\right)^{z}\right>$ appears as a coefficient of the Dzyaloshinsky--Moriya interaction in the Hamiltonian Eq.~(\ref{eqn1}), while $\textrm{Re}\left<S_{j}^{+}S_{i}^{-}\right>=\left<S_{j}^{x}S_{i}^{x}+S_{j}^{y}S_{i}^{y}\right>$ is connected to the Heisenberg interaction. Therefore, it is expected that the ratio of the real and imaginary parts of the correlation function follow the ratio of the interaction coefficients which they  are attributed to,
\begin{eqnarray}
\frac{D_{ij}}{J_{ij}}\approx\frac{\textrm{Im}\left<S_{j}^{+}S_{i}^{-}\right>}{\textrm{Re}\left<S_{j}^{+}S_{i}^{-}\right>}.\label{eqn21a}
\end{eqnarray}

Substituting Eq.~(\ref{eqn21a}) into Eqs.~(\ref{eqn10})-(\ref{eqn11}) yields a very similar temperature dependence for $\mathscr{A}$ and $\mathscr{D}$. This is in agreement with the observation that the wavelength of spin spirals ($\lambda\propto\mathscr{A}/\mathscr{D}$) is practically independent of the temperature\cite{Hasselberg}.
%This property implies the relation $\frac{D_{ij}}{J_{ij}}\approx\frac{\textrm{Im}\left<S_{j}^{+}S_{i}^{-}\right>}{\textrm{Re}\left<S_{j}^{+}S_{i}^{-}\right>}$, and a very similar temperature dependence for $\mathscr{A}$ and $\mathscr{D}$. For an exact relation in the case of only nearest-neighbor interactions see Appendix~\ref{secS3}.

Finally, the Dzyaloshinsky--Moriya interaction also influences the temperature dependence of the anisotropy term $\mathscr{K}^{zz}$ as shown in Eq.~(\ref{eqn12}). This is surprising because it is known that the Dzyaloshinsky--Moriya interaction does not influence the spin wave spectrum of ferromagnetic systems at $\boldsymbol{k}=\boldsymbol{0}$ at zero temperature, it only induces an asymmetry between $\boldsymbol{k}$ and $-\boldsymbol{k}$\cite{Udvardi}. Since the system gains energy from the Dzyaloshinsky--Moriya interaction in Eq.~(\ref{eqn1}) if $D_{ij}\textrm{Im}\left<S_{j}^{+}S_{i}^{-}\right>$ is positive, the anisotropy term induced by the Dzyaloshinsky--Moriya interaction always has a positive sign and increases at low temperature with the fluctuations. This is contrary to the temperature dependence of the single-ion anisotropy, which always decreases ($\mathscr{K}^{zz}\propto m^{3}$ at low temperature). While all the spins are parallel in the ferromagnetic ground state, at finite temperature the spins are fluctuating, and the Dzyaloshinsky--Moriya interaction opens a finite average angle $\textrm{Im}\left<S_{j}^{+}S_{i}^{-}\right>\propto\sin\vartheta$ between them, which induces an anisotropy term in the spin wave spectrum. An analogy can be drawn between the correlated random fluctuations of the spins and the similar anisotropy effect observed in spin glasses\cite{Fert2}, where the finite average angle between the spins appears because of the random relative positions of the magnetic atoms and nonmagnetic impurities\cite{Levy}.% as shown in Eq.~(\ref{eqn5}).

\section{Comparison to Monte Carlo simulations\label{sec3}}

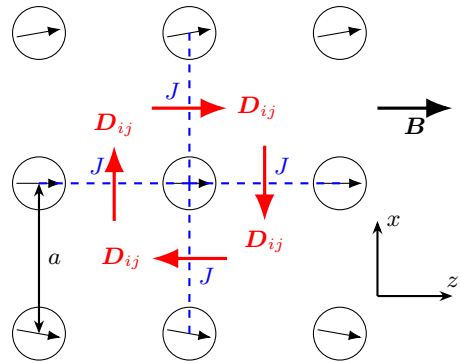
\begin{figure}
\centering
\begin{tikzpicture}
\foreach \x in {-2,0,2}
{
\foreach \y in {-2,0,2}
{
\node at (\x,\y) [draw,circle,minimum size = 20] {};
\draw[black,-{Latex[length=5]}] (\x,\y)+(180+5*\y:0.3) -- +(0+5*\y:0.3);
}
}
\draw[red,-{Latex[length=10]},very thick]  (1,0.5) -- (1,-0.5) node[below] {$\boldsymbol{D}_{ij}$};
\draw[red,-{Latex[length=10]},very thick]  (-1,-0.5)  -- (-1,0.5) node[above] {$\boldsymbol{D}_{ij}$};
\draw[red,-{Latex[length=10]},very thick]  (0.5,-1) -- (-0.5,-1) node[left] {$\boldsymbol{D}_{ij}$};
\draw[red,-{Latex[length=10]},very thick]  (-0.5,1)  -- (0.5,1) node[right] {$\boldsymbol{D}_{ij}$};
\draw[black,{Stealth[length=5]}-{Stealth[length=5]},thick] (-2,-2) -- node[right] {$a$} (-2,0) ;
\draw[blue,dashed,thick]  (-2,0) -- node[above left] {$J$} (0,0) ;
\draw[blue,dashed,thick]  (0,0) -- node[above right] {$J$} (2,0) ;
\draw[blue,dashed,thick]  (0,-2) -- node[below right] {$J$} (0,0) ;
\draw[blue,dashed,thick]  (0,0) -- node[above left] {$J$} (0,2) ;
\draw[black,-{Latex[length=10]}, very thick] (2.5,1) -- node[below] {$\boldsymbol{B}$} (3.5,1) ;
\draw[black,-{Stealth[length=5]},thick] (2.5,-1.5) -- (3.5,-1.5)  node[above] {$z$};
\draw[black,-{Stealth[length=5]},thick] (2.5,-1.5) -- (2.5,-0.5)  node[right] {$x$};
\end{tikzpicture}
\caption{Sketch of the square lattice and the interaction parameters used for the calculations and the Monte Carlo simulations. 
% The dimensionless interaction parameters are $J=1, \left|\boldsymbol{D}_{ij}\right|=D=0.2, B^{z}=0.1, \mu_{s}=1$. 
The considered spin waves are propagating along the $x$ direction.\label{fig1}}
\end{figure}

In order to illustrate the theory outlined above for obtaining the effective temperature-dependent interaction coefficients Eqs.~(\ref{eqn10})-(\ref{eqn12}), as a model system we considered a square lattice, representing a magnetic monolayer on a cubic $(001)$ surface. We chose an ultrathin magnetic film in order to examine the pronounced role of the spin fluctuations, and because such systems play an important role in suggested applications based on the Dzyaloshinsky--Moriya interaction\cite{Fert,Garcia-Sanchez}. As displayed in Fig.~\ref{fig1}, we used only nearest-neighbor Heisenberg $J$ and Dzyaloshinsky--Moriya $D$ interactions, with the Dzyaloshinsky--Moriya vectors pointing perpendicular to the lattice vectors due to the $C_{4\textrm{v}}$ symmetry of the system. The second-order single-ion anisotropy in Eq.~(\ref{eqn1}) can only describe an out-of-plane easy axis or an easy plane in the considered system by symmetry, and this term was neglected to simplify the calculations. We applied an in-plane magnetic field $\boldsymbol{B}$ along the $z$ direction to force the system in an in-plane ferromagnetic state and calculated the spin wave frequencies with propagation vectors along the perpendicular $x$ direction. This is a standard procedure for the experimental determination of the Dzyaloshinsky--Moriya interaction in ultrathin films, for example see Refs.~\cite{Zakeri,Nembach,Belmeguenai}.

In the present case, the spin wave spectrum Eq.~(\ref{eqn9}) along the $x$ direction may be expressed as
\begin{eqnarray}
\frac{\mu_{s}}{\gamma}\omega_{\boldsymbol{k}}\left(T\right)=&&2\mathcal{J}\left(1-\cos\left(k^{x}a\right)\right)+2\mathcal{D}\sin\left(k^{x}a\right)\nonumber
\\
&&+2\mathcal{K}^{zz}+\mu_{s}B^{z},\label{eqn22}
\end{eqnarray}
with the connection to the atomistic parameters $J$ and $D$ as defined in Eqs.~(\ref{eqn10})-(\ref{eqn12}). Importantly, the anisotropy term $\mathcal{K}^{zz}$ is only induced by the presence of thermal fluctuations and the Dzyaloshinsky--Moriya interactions. In the considered system with only nearest-neighbor interactions, the ratio $\mathcal{D}/\mathcal{J}$ is independent of the temperature, emphasizing the strong analogy between the Heisenberg and Dzyaloshinsky--Moriya exchange interactions. For a proof see Appendix~\ref{secS3}.
%\begin{eqnarray}
%\mathcal{J}^{x,y}&=&mJ+m^{2}J\:\textrm{Re}I^{x,y},\label{eqn23}
%\\
%\mathcal{D}&=&mD+m^{2}J\:\textrm{Im}I^{y},\label{eqn24}
%\\
%\mathcal{K}^{xx}&=&D\:\textrm{Im}I^{y}.\label{eqn25}
%\end{eqnarray}

%The expressions $I^{x,y}$ appearing in Eqs.~(\ref{eqn23})-(\ref{eqn25}) for an infinite lattice are given by
%\begin{eqnarray}
%I^{x,y}=\left(\frac{a}{2\pi}\right)^{2}\int_{\textrm{BZ}}\textrm{e}^{-\textrm{i}k^{x,y}a}\frac{\gamma}{\mu_{s}}\frac{k_{\textrm{B}}T}{\omega_{\boldsymbol{k}}\left(T\right)}\textrm{d}^{2}\boldsymbol{k}.\label{eqn26}
%\end{eqnarray}

%At zero temperature, Eqs.~(\ref{eqn23})-(\ref{eqn25}) simplify to the microscopic interaction parameters, which satisfy the $C_{4\textrm{v}}$ symmetry of the system with $\mathcal{J}^{x}=\mathcal{J}^{y}=J$ and $\mathcal{K}^{xx}=0$. However, applying a magnetic field along the $x$ direction selects a preferred orientation of the spins and breaks this symmetry, which leads to $\mathcal{J}^{x}\neq\mathcal{J}^{y}$ and $\mathcal{K}^{xx}\neq 0$ in the presence of a finite Dzyaloshinsky--Moriya interaction. $B^{x}$ and $D$ do not break the $\omega_{k^{x},k^{y}}=\omega_{-k^{x},k^{y}}$ symmetry of the spectrum, which implies that $I^{x}$ will remain real at all temperatures. For this system, it can be calculated analytically that $\frac{\mathcal{J}^{y}}{J}=\frac{\mathcal{D}}{D}$ will always hold, meaning that the temperature dependence of the Heisenberg and the Dzyaloshinsky--Moriya exchange interactions is exactly the same within this model.

\begin{figure}
\centering
\includegraphics[width=\columnwidth]{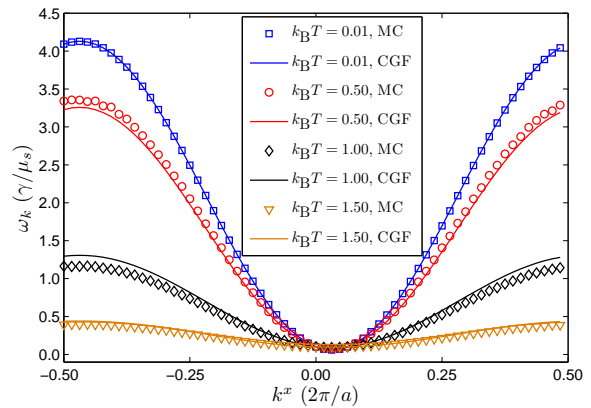}
\caption{Calculated spin wave spectrum from the Monte Carlo simulations (MC) and from Green's function theory in Callen's formulation (CGF). The dimensionless interaction parameters are $J=1, D=-0.2, B^{z}=0.1, \mu_{s}=1$.%{\disp Comment unai: $T_{c}$ in terms of $k_{\textrm{B}}T$?} 
\label{figS1}}
\end{figure}

For checking the accuracy of the theoretical model, we performed Monte Carlo simulations on an $N=64\times64$ lattice. For the details of the simulations see Appendix~\ref{secS4}. The simulations converge to the thermal equilibrium of the system described by the Hamiltonian Eq.~(\ref{eqn1}), and include all higher-order correlation functions neglected in the model. The spin wave frequencies may be expressed by rewriting Eq.~(\ref{eqn15}) in the form
\begin{eqnarray}
\omega_{\boldsymbol{k}}\left(T\right)=\frac{\gamma}{\mu_{s}}k_{\textrm{B}}T\frac{2m}{N\left<S^{-}_{\boldsymbol{k}}S^{+}_{-\boldsymbol{k}}\right>},\label{eqn27}
\end{eqnarray}
where the right-hand side contains only expectation values in thermal equilibrium, which can be calculated from Monte Carlo simulations. The softening of the spin wave frequencies with temperature and the comparison to the theoretical model is illustrated in Fig.~\ref{figS1}.

\begin{figure}
\centering
\includegraphics[width=\columnwidth]{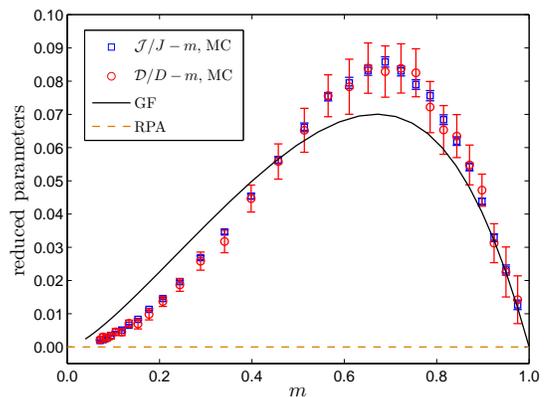}
\caption{%Effective temperature-dependent coefficients as a function of magnetization, with each parameter divided by its zero-temperature value.
Fluctuation corrections to the effective temperature-dependent coefficients as a function of magnetization: comparison between the results of Monte Carlo simulations (MC) and Green's function theory in Callen's formulation (CGF). These corrections are not included in the random phase approximation (RPA). The theoretical models give the same prediction for the Heisenberg and Dzyaloshinsky--Moriya terms. The error bars indicate the uncertainty of the fitting parameters. The dimensionless interaction parameters are $J=1, D=-0.2, B^{z}=0.1, \mu_{s}=1$.\label{fig2}}
\end{figure}

After obtaining the spin wave frequencies, an expression of the form of Eq.~(\ref{eqn22}) was fitted to the dispersion relation to obtain the effective finite-temperature coupling coefficients. The results of the Monte Carlo simulations are compared to the theoretical predictions in Fig.~\ref{fig2}. The fluctuation corrections can modify the interactions significantly, by almost 10\% of their zero-temperature value at maximum. The deviation from the result of the random phase approximation or mean-field model due to spin correlations is most pronounced at low temperature. Including the correlation corrections using Green's function theory in Callen's formulation as discussed above gives reasonable quantitative agreement with the simulation results over the whole range in magnetization, while higher-order corrections are less significant. The simulations also confirm that the temperature dependence of the Heisenberg and Dzyaloshinsky--Moriya interactions is very similar; the two functions completely coincide in the theoretical model as discussed above.

For a quantitative comparison between theory and simulations, we transformed the temperature-dependent coefficients to the corresponding quantities in the micromagnetic model -- see Eqs.~(\ref{eqn19})-(\ref{eqn20}) --, and fitted power functions of the form $\mathscr{A},\mathscr{D}\propto m^{\kappa_{\mathscr{A},\mathscr{D}}}$ to the data in the range $0.9\le m\le1.0$. In the case of the simulation results, we obtain $\kappa_{\mathscr{A}}=\kappa_{\mathscr{D}}=1.54$ (the values agree within the given precision), Green's function theory in Callen's formulation yields $\kappa_{\textrm{\tiny CGF}}=1.57$, while the random phase approximation leads to the well-known mean-field result $\kappa_{\textrm{\tiny RPA}}=2$. Exponents between $1.66$ and $1.76$ were calculated for the temperature-dependence of the Heisenberg exchange stiffness of three-dimensional magnets in Ref.~\cite{Atxitia}, indicating that the fluctuation corrections play an even more pronounced role in the presently considered two-dimensional ultrathin film. Note that as the magnetization decreases with increasing temperature, the correlation corrections also tend to zero, and the mean-field exponent $\kappa=2$ is recovered both from simulations and theory.
%, but it converges to zero as the magnetization decreases with increasing temperature. Approaching the critical temperature $T_{c}$, the temperature dependence of the interaction parameters tends to the result of the mean-field approximation, %which is correct near the critical temperature. This means 
%$m\sim(T- T_{c})^{1/2}$ near $T_{c}$, leading to a linear dependence of the Heisenberg and Dzyaloshinsky--Moriya exchange on temperature, in agreement with the Landau theory of phase transitions.
%; at high temperature all interaction parameters scale with the magnetization.
%It can also be seen that the temperature dependence of the Heisenberg and Dzyaloshinsky--Moriya interactions is very similar in the simulations, in agreement with the theoretical prediction.

\begin{figure}
\centering
\includegraphics[width=\columnwidth]{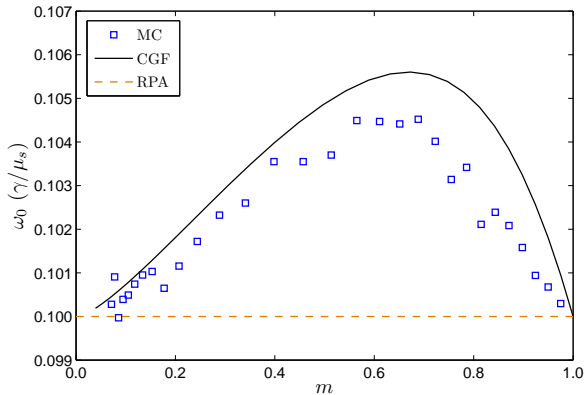}
\caption{Energy of the spin wave at wave vector $\boldsymbol{k}=\boldsymbol{0}$ as a function of magnetization: comparison between results of Monte Carlo simulations (MC), Green's function theory in Callen's formulation (CGF) and in the random phase approximation (RPA). The dimensionless interaction parameters are $J=1, D=-0.2, B^{x}=0.1, \mu_{s}=1$.\label{fig3}}
\end{figure}

Another important prediction of the theoretical model is the anisotropy term induced by the Dzyaloshinsky--Moriya interaction at finite temperature. This can be visualized by calculating the frequency of the spin wave with zero wave vector, $\frac{\mu_{s}}{\gamma}\omega_{\boldsymbol{0}}=2\mathcal{K}^{zz}+\mu_{s}B$, as shown in Fig.~\ref{fig3}. In the considered system, for the induced anisotropy Green's function theory predicts
\begin{eqnarray}
\mathcal{K}^{zz}=\frac{D^{2}}{J}\left(\frac{\mathcal{J}}{J}-m\right).\label{eqn27a}
\end{eqnarray}

The energy gain from spin canting $D^{2}/J$ is generally several percents of the Heisenberg exchange $J$ in ultrathin films, and the fluctuation corrections can almost reach a maximum of 10\% as shown in Fig.~\ref{fig2}. Since the typical strength of the Heisenberg exchange is on the order of $10\,\textrm{meV}$, the maximum of the induced anisotropy may be on the order of $0.1\,\textrm{meV}$, which is comparable in magnitude to the demagnetization anisotropy induced by dipolar interactions in ferromagnetic monolayers -- for typical parameter values obtained from \textit{ab initio} calculations see e.g. Ref.~\cite{Vida}.

%{\disp Comment Unai: I would point out that the $m$ dependence of all the corrections have a similar shape since they just come out from $ \textrm{Im}\left<S_{j}^{+}S_{i}^{-}\right>$}
% The figure demonstrates that the effect is also observable in the simulations, although with a slightly lower magnitude than in the theoretical model.

%\begin{figure}
%\centering
%\includegraphics[width=\columnwidth]{K0D02B01gap.eps}
%\caption{Energy of the spin wave at wave vector $\boldsymbol{k}=\boldsymbol{0}$ as a function of temperature. The figure shows the comparison between the results of Monte Carlo simulations (MC), Green's function theory (GF) and the random phase approximation (RPA). The dimensionless interaction parameters are $J=-1, \left|\boldsymbol{D}_{ij}\right|=D=0.2, B=0.1, \mu_{s}=1$.\label{fig3}}
%\end{figure}

%Figure~\ref{fig3} demonstrates that in the presence of the Dzyaloshinsky--Moriya interaction, the energy of the magnon in the center of the Brillouin zone increases with temperature, leading to the appearance of an effective anisotropy term. Although Green's function theory correctly predicts this effect, quantitatively it significantly overestimates it in the present system.

\section{Summary}

In summary, we established a connection between the temperature-independent atomistic interaction parameters $J_{ij},\boldsymbol{D}_{ij},K^{zz}$ in Eq.~(\ref{eqn1}) and the effective temperature-dependent micromagnetic interaction parameters $\mathscr{A},\mathscr{D},\mathscr{K}^{zz}$ in Eq.~(\ref{eqn16}) by calculating the spin wave spectrum. In the atomistic calculations, we relied on the classical version of the Green's function formalism as formulated in Ref.~\cite{Callen}. By comparing the theoretical calculations to Monte Carlo simulations, we demonstrated on a simple model system that the method describes the finite-temperature corrections due to spin fluctuations with a high precision, while the well-known mean-field or random phase approximation gives significantly less accurate results. The correlation corrections for the Heisenberg and Dzyaloshinsky--Moriya exchange interactions are very similar and are determined by the spatial decay of the transversal spin correlation function, while the single-ion anisotropy term must be treated differently. We also showed that the presence of the Dzyaloshinsky--Moriya interaction can give rise to an additional anisotropy accompanied by an increasing spin wave frequency at zero wave vector with increasing temperature, although it does not modify it at zero temperature. This effect can be attributed to the Dzyaloshinsky--Moriya interactions in connection with the finite angle between the fluctuating transversal spin components of the spins.
%, analogously how they introduce a finite angle due to the random relative positions of atoms in spin glasses\cite{Levy}. 

Overall, it can be concluded that the method presented here may be used for the determination of temperature-dependent micromagnetic interaction parameters in multiscale models, possibly circumventing time-consuming atomistic spin dynamics simulations. Due to the simple form of Eqs.~(\ref{eqn10})-(\ref{eqn12}), the close analogy with quantum spin models, the fact that the expressions do not explicitly rely on the symmetry or dimension of the system, and the possible generalization to other types of magnetic order, it is expected that the results presented in this paper may motivate further studies for the determination of temperature-dependent micromagnetic parameters in magnetic materials.

\begin{acknowledgments}

The authors would like to thank Rocio Yanes for enlightening discussions. Financial support for this work was provided by the Deutsche Forschungsgemeinschaft via SFB 767, SFB 668, and the Priority Program SpinCaT, and by the National Research, Development and Innovation Office of Hungary under project No. K115575.

\end{acknowledgments}

\appendix

\section{Green's function theory\label{secS1}}

Here it will be discussed how the excitation frequencies of the classical spin system Eq.~(\ref{eqn1}) may be calculated at finite temperature. Following the derivation for the quantum case\cite{Callen}, first one has to define the Poisson brackets of the spin components\cite{Bastardis}
\begin{eqnarray}
\left\{S_{i}^{\alpha},S_{j}^{\beta}\right\}=-\frac{\gamma}{\mu_{s}}\varepsilon^{\alpha\beta\gamma}\delta_{ij}S_{i}^{\gamma},\label{eqnS1}
\end{eqnarray}
which generate the equation of motion through the well-known formula
\begin{eqnarray}
\partial_{t}S_{i}^{\alpha}=\left\{S_{i}^{\alpha},H\right\}=\frac{\gamma}{\mu_{s}}\varepsilon^{\alpha\beta\gamma}S_{i}^{\alpha}\frac{\partial H}{\partial S_{i}^{\beta}}.\label{eqnS2}
\end{eqnarray}

If the ground state is ferromagnetic along the $z$ direction as supposed in the main part of the manuscript, it is advised to introduce the variables $S_{i}^{\pm}=S_{i}^{x}\pm S_{i}^{y}$ satisfying
\begin{eqnarray}
\left\{S_{i}^{z},S_{j}^{\pm}\right\}&=&\pm\textrm{i}\frac{\gamma}{\mu_{s}}\delta_{ij}S_{i}^{\pm},\label{eqnS3}
\\
\left\{S_{i}^{+},S_{j}^{-}\right\}&=&2\textrm{i}\frac{\gamma}{\mu_{s}}\delta_{ij}S_{i}^{z}.\label{eqnS4}
\end{eqnarray}

Using the transformed variables, the Hamiltonian Eq.~(\ref{eqn1}) may be rewritten as
\begin{align}
H=&-\frac{1}{2}\sum_{i,j}\Big(J_{ij}S_{i}^{z}S_{j}^{z}+\textrm{Re}\left[\left(J_{ij}+\textrm{i}D_{ij}^{z}\right)S_{i}^{+}S_{j}^{-}\right]\Big)\nonumber
\\
&-\sum_{i}K^{zz}\left(S_{i}^{z}\right)^{2}-\mu_{s}\sum_{i}B^{z}S_{i}^{z}\nonumber
\\
&-\frac{1}{2}\sum_{i,j}\Big(S_{i}^{z}\textrm{Re}\left[\left(D_{ij}^{y}+\textrm{i}D_{ij}^{x}\right)S_{j}^{+}\right]\nonumber
\\
&-\textrm{Re}\left[\left(D_{ij}^{y}+\textrm{i}D_{ij}^{x}\right)S_{i}^{+}\right]S_{j}^{z}\Big).\label{eqnS5}
\end{align}

In spin wave theory, the equations of motion must be linearized in the variables $S_{i}^{\pm}\ll 1$. It can be shown that the terms proportional to $D_{ij}^{x}$ and $D_{ij}^{y}$ only yield higher-order corrections, and they will be neglected in further calculations. Consequently, we will also drop the $z$ index of the Dzyaloshinsky--Moriya vector component parallel to the magnetization. The equations for $S_{i}^{+}$ and $S_{i}^{-}$ decouple, and may be diagonalized in Fourier space,
\begin{eqnarray}
S^{-}_{\boldsymbol{k}}=\frac{1}{\sqrt{N}}\sum_{i}\textrm{e}^{-\textrm{i}\boldsymbol{k}\boldsymbol{R}_{i}}S_{i}^{-},\label{eqnS6}
\end{eqnarray}
yielding
\begin{eqnarray}
\partial_{t}S_{\boldsymbol{k}}^{-}=-\textrm{i}\omega_{\boldsymbol{k}}^{T=0}S_{\boldsymbol{k}}^{-},\label{eqnS7}
\end{eqnarray}
with the spin wave frequencies
\begin{eqnarray}
\omega_{\boldsymbol{k}}^{T=0}=\frac{\gamma}{\mu_{s}}\left(J_{\boldsymbol{0}}-J_{\boldsymbol{k}}-\textrm{i}D_{\boldsymbol{k}}+2K^{zz}+\mu_{s}B^{z}\right).\label{eqnS8}
\end{eqnarray}

The Fourier transforms of the interaction coefficients are defined as
\begin{eqnarray}
J_{\boldsymbol{k}}&=&\sum_{\boldsymbol{R}_{i}-\boldsymbol{R}_{j}}\textrm{e}^{-\textrm{i}\boldsymbol{k}\left(\boldsymbol{R}_{i}-\boldsymbol{R}_{j}\right)}J_{ij},\label{eqnS9}
\\
\textrm{i}D_{\boldsymbol{k}}&=&\sum_{\boldsymbol{R}_{i}-\boldsymbol{R}_{j}}\textrm{e}^{-\textrm{i}\boldsymbol{k}\left(\boldsymbol{R}_{i}-\boldsymbol{R}_{j}\right)}\textrm{i}D_{ij}^{z}.\label{eqnS10}
\end{eqnarray}

Note that Eq.~(\ref{eqnS10}) is real-valued, because $\boldsymbol{D}_{ij}$ is antisymmetric in the lattice indices.

At finite temperature, following Refs.~\cite{Callen,Bastardis} we will consider the time-dependent Green's function
\begin{eqnarray}
G_{ij}\left(t;r\right)=\theta\left(t\right)\left<\left\{S_{i}^{-}\left(t\right),\textrm{e}^{rS_{j}^{z}\left(0\right)}S_{j}^{+}\left(0\right)\right\}\right>,\label{eqnS11}
\end{eqnarray}
where $\theta\left(t\right)$ is the Heaviside function, $\left<\right>$ denotes averaging in thermal equilibrium, and $r$ is a real parameter. Instead of the homogeneous Eq.~(\ref{eqnS2}), this satisfies the inhomogeneous equation of motion
\begin{eqnarray}
\partial_{t}G_{ij}=&&\delta\left(t\right)\left<\left\{S_{i}^{-},\textrm{e}^{rS_{j}^{z}}S_{j}^{+}\right\}\right>\nonumber
\\
&&+\theta\left(t\right)\!\left<\!\left\{\!\left\{S_{i}^{-}\left(t\right),H\right\},\textrm{e}^{rS_{j}^{z}\left(0\right)}S_{j}^{+}\left(0\right)\!\right\}\!\right>\!\!.\label{eqnS12}
\end{eqnarray}

The Poisson bracket
\begin{eqnarray}
\left\{S_{i}^{-}\left(t\right),H\right\}=&&\textrm{i}\frac{\gamma}{\mu_{s}}\Big[\sum_{l}\left(-J_{il}S_{i}^{-}S_{l}^{z}+\left(J_{il}+\textrm{i}D_{il}\right)S_{i}^{z}S_{j}^{-}\right)\nonumber
\\
&&-2K^{zz}S_{i}^{z}S_{i}^{-}-\mu_{s}B^{z}S_{i}^{-}\Big]\label{eqnS13}
\end{eqnarray}
introduces higher-order Green's functions on the right-hand side of Eq.~(\ref{eqnS12}). These are handled within the decoupling approximation,
\begin{align}
\theta\left(t\right)\left<\left\{S_{i}^{-}\left(t\right)S_{l}^{z}\left(t\right),\textrm{e}^{rS_{j}^{z}\left(0\right)}S_{j}^{+}\left(0\right)\right\}\right>\approx\nonumber
\\
\approx mG_{ij}-\alpha\left<S_{i}^{-}S_{l}^{+}\right>G_{lj}.\label{eqnS14}
\end{align}

It was demonstrated in Ref.~\cite{Bastardis} that in the classical limit Eq.~(\ref{eqnS14}) may be used both for exchange interactions and single-ion anisotropy terms. In the literature there exist several schemes for the decoupling coefficient $\alpha$; we used the value $\alpha=\frac{m}{2}$ from Refs.~\cite{Callen,Bastardis} in the main text, since we found that generally this gives the best agreement with the spin wave spectrum calculated from the simulations.

After performing Fourier transformation in time ($\partial_{t}\rightarrow -\textrm{i}\omega$) and space,
\begin{eqnarray}
G_{\boldsymbol{k}}&=&\frac{1}{N}\sum_{\boldsymbol{R}_{i}-\boldsymbol{R}_{j}}\textrm{e}^{-\textrm{i}\boldsymbol{k}\left(\boldsymbol{R}_{i}-\boldsymbol{R}_{j}\right)}G_{ij},\label{eqnS15}
\\
\left<S_{\boldsymbol{k}}^{-}S_{-\boldsymbol{k}}^{+}\right>&=&\frac{1}{N}\sum_{\boldsymbol{R}_{i}-\boldsymbol{R}_{j}}\textrm{e}^{-\textrm{i}\boldsymbol{k}\left(\boldsymbol{R}_{i}-\boldsymbol{R}_{j}\right)}\left<S_{i}^{-}S_{j}^{+}\right>,\label{eqnS16}
\end{eqnarray}
Eq.~(\ref{eqnS12}) may be rewritten as
\begin{eqnarray}
\left(\omega-\omega_{\boldsymbol{k}}\right)G_{\boldsymbol{k}}=\frac{1}{2\pi}\frac{\gamma}{\mu_{s}}\frac{1}{N}\Theta\left(r\right),\label{eqnS17}
\end{eqnarray}
with
\begin{eqnarray}
\Theta\left(r\right)=\frac{\textrm{i}\mu_{s}}{\gamma}\left<\left\{S_{i}^{-},\textrm{e}^{rS_{j}^{z}}S_{j}^{+}\right\}\right>.\label{eqnS18}
\end{eqnarray}

Note that $\Theta\left(0\right)=2m$ from the Poisson bracket Eq.~(\ref{eqnS4}).

The spin wave frequencies read
\begin{align}
\omega_{\boldsymbol{k}}=&\frac{\gamma}{\mu_{s}}\Big[\left(J_{\boldsymbol{0}}-J_{\boldsymbol{k}}-\textrm{i}D_{\boldsymbol{k}}+2K^{zz}\right)m+\mu_{s}B^{z}\nonumber
\\
&-\alpha\sum_{\boldsymbol{k}'}\left(J_{\boldsymbol{k}-\boldsymbol{k}'}-J_{\boldsymbol{k}'}-\textrm{i}D_{\boldsymbol{k}'}+2K^{zz}\right)\left<S_{\boldsymbol{k}'}^{-}S_{-\boldsymbol{k}'}^{+}\right>\Big].\label{eqnS19}
\end{align}

By performing inverse Fourier transformation in space, one arrives at the expression
\begin{eqnarray}
\omega_{ij}=&&\frac{\gamma}{\mu_{s}}\bigg(-\left(J_{ij}+\textrm{i}D_{ij}\right)m-\alpha J_{ij}\left<S_{j}^{+}S_{i}^{-}\right>\nonumber
\\
&&+\delta_{ij}\bigg[\left(\sum_{l}J_{il}+2K^{zz}\right)m+\alpha\sum_{l}\left(J_{il}+\textrm{i}D_{il}\right)\nonumber
\\
&&\times\left<S_{l}^{+}S_{i}^{-}\right>-\alpha2K^{zz}\left<S_{i}^{+}S_{i}^{-}\right>\bigg]\bigg).\label{eqnS20}
\end{eqnarray}

Collecting the real and imaginary parts in the off-diagonal ($i\!\!\!\!\!\neq\!\!\!\!\!j$) part of Eq.~(\ref{eqnS20}) yields Eqs.~(\ref{eqn10})-(\ref{eqn11}) for the effective temperature-dependent Heisenberg and Dzyaloshinsky--Moriya interactions, while the extra terms in the diagonal part may be collected into the anisotropy term Eq.~(\ref{eqn12}).

In order to solve Eqs.~(\ref{eqnS17})-(\ref{eqnS19}), one has to introduce the spectral density
\begin{eqnarray}
S_{ij}\left(\omega\right)\!=\!\frac{\textrm{i}}{2\pi}\lim_{\delta\rightarrow 0}\left(G_{ij}\left(\omega+\textrm{i}\delta\right)-G_{ij}\left(\omega-\textrm{i}\delta\right)\right)\!,\label{eqnS21}
\end{eqnarray}
and use the spectral theorem
\begin{eqnarray}
\left<\textrm{e}^{rS_{j}^{z}}S_{j}^{+}S_{i}^{-}\right>=\int_{-\infty}^{\infty} \frac{k_{\textrm{B}}T}{\omega}S_{ij}\left(\omega\right)\textrm{d}\omega,\label{eqnS22}
\end{eqnarray}
the classical limit of the corresponding quantum expression\cite{Frobrich}. Since Eq.~(\ref{eqnS17}) describes pure single-particle excitations in the current approximation, it simplifies to Eq.~(\ref{eqn15}) for $r=0$ in Fourier space.

In real space, the appropriate form of Eq.~(\ref{eqnS22}) is
\begin{eqnarray}
\left<\textrm{e}^{rS_{i}^{z}}S_{i}^{+}S_{i}^{-}\right>=\Phi\Theta\left(r\right),\label{eqnS23}
\end{eqnarray}
with $\Phi$ from Eq.~(\ref{eqn14}). Both sides of Eq.~(\ref{eqnS23}) may be expressed by the momentum generating function
\begin{eqnarray}
\Omega\left(r\right)=\left<\textrm{e}^{rS_{i}^{z}}\right>\label{eqnS24}
\end{eqnarray}
by using the Poisson brackets, yielding the differential equation
\begin{eqnarray}
\Omega''+\frac{2\Phi}{1+r\Phi}\Omega'-\Omega=0,\label{eqnS25}
\end{eqnarray}
where $'$ denotes differentiation with respect to $r$.

The solution of Eq.~(\ref{eqnS25}) satisfying $\Omega\left(0\right)=1$ and regularity conditions for $\Omega$ reads
% the criterion that $\Omega$ must remain finite for finite $r$ reads
\begin{eqnarray}
\Omega\left(r\right)=\frac{1}{1+r\Phi}\frac{\sinh\left(r+\frac{1}{\Phi}\right)}{\sinh\frac{1}{\Phi}}.\label{eqnS26}
\end{eqnarray}
Equation~(\ref{eqnS26}) may also be obtained as the classical limit of the corresponding quantum expression\cite{Callen}, or by calculating $\left<\textrm{e}^{rS_{i}^{z}}\right>$ in a mean-field model and using the analogy $\frac{1}{\Phi}\leftrightarrow\frac{\mu_{s}B_{\textrm{MF}}}{k_{\textrm{B}}T}$ between the probability densities of mean-field and Green's function theories\cite{Callen3}. Finally, we mention that calculating $\Omega'\left(0\right)=m$ yields Eq.~(\ref{eqn13}) for the magnetization.

\section{Micromagnetic model\label{secS2}}

In the general case, the micromagnetic free energy functional of a three-dimensional system described by the atomistic Hamiltonian Eq.~(\ref{eqn1}) is given by
\begin{eqnarray}
F=&&\int\sum_{\alpha,\beta,\gamma}\partial_{\alpha}S^{\gamma}\mathscr{A}^{\alpha\beta}\partial_{\beta}S^{\gamma}+\sum_{\alpha\beta}\mathscr{D}^{\alpha\beta}L^{\alpha\beta}\left(\boldsymbol{S}\right)\nonumber
\\
&&-\mathscr{K}^{zz}\left(S^{z}\right)^{2}-\mathscr{M}B^{z}S^{z}\textrm{d}^{3}\boldsymbol{r},\label{eqnS27}
\end{eqnarray}
where $\mathscr{A}^{\alpha\beta}$ and $\mathscr{D}^{\alpha\beta}$ are both $3\times 3$ tensors\cite{Schweflinghaus}. %Equation~(\ref{eqn16}) in the main text corresponds to the density of $F$ per unit area perpendicular to the $x$ direction, along which direction the spin modulations occur.

For the Heisenberg exchange interaction, Eq.~(\ref{eqnS27}) expresses that the $J_{ij}$ coefficients may be anisotropic in real space, i.e. they may differ for neighbors along different directions. However, all anisotropy in spin space is included in the $\mathscr{K}^{zz}$ term, since the model did not contain two-ion anisotropy terms. The Heisenberg exchange tensor may be expressed as
\begin{eqnarray}
\mathscr{A}^{\alpha\beta}=&&\frac{1}{4}\frac{m}{\upsilon_{\textrm{\tiny WS}}}\sum_{\boldsymbol{R}_{i}-\boldsymbol{R}_{j}}\mathcal{J}_{ij}\left(R^{\alpha}_{j}-R^{\alpha}_{i}\right)\left(R^{\beta}_{j}-R^{\beta}_{i}\right),\label{eqnS28}
\end{eqnarray}
indicating that it is symmetric in the Cartesian indices.

For the Dzyaloshinsky--Moriya interaction, the linear Lifshitz invariant is usually defined in the form
\begin{eqnarray}
L^{(\beta)}_{\alpha\gamma}=S^{\alpha}\partial_{\beta}S^{\gamma}-S^{\gamma}\partial_{\beta}S^{\alpha},\label{eqnS29}
\end{eqnarray}
which is antisymmetric in the $\alpha$ and $\gamma$ indices. Due to considering three-dimensional spins, these two indices may simply be replaced by the perpendicular direction
\begin{eqnarray}
L^{\alpha\beta}=\frac{1}{2}\sum_{\gamma,\delta}\varepsilon^{\alpha\gamma\delta}L_{\gamma\delta}^{(\beta)},\label{eqnS30}
\end{eqnarray}
the tensorial notation used in Eq.~(\ref{eqnS28})\cite{Schweflinghaus}. The corresponding Dzyaloshinsky--Moriya tensor reads
\begin{eqnarray}
\mathscr{D}^{\alpha\beta}=-\frac{m}{2\upsilon_{\textrm{\tiny WS}}}\sum_{\boldsymbol{R}_{i}-\boldsymbol{R}_{j}}\mathcal{D}^{\alpha}_{ij}\left(R^{\beta}_{j}-R^{\beta}_{i}\right).\label{eqnS31}
\end{eqnarray}

Equation~(\ref{eqnS31}) has no specific symmetry properties, since the first index describes the rotational plane of the spins, while the second stands for the direction of the modulation. For spin waves, the rotational plane is perpendicular to the ferromagnetic direction. If we denote the ferromagnetic direction by $\boldsymbol{e}_{\textrm{FM}}$ and calculate the spin wave frequencies along the $\boldsymbol{e}_{\boldsymbol{k}}$ direction, only a single component of the Dzyaloshinsky--Moriya tensor may be calculated (cf. Eq.~(\ref{eqn16})),
\begin{eqnarray}
\mathscr{D}=\sum_{\alpha,\beta}e^{\alpha}_{\textrm{FM}}\mathscr{D}^{\alpha\beta}e^{\beta}_{\boldsymbol{k}}.\label{eqnS32}
\end{eqnarray}

The symmetry of the system determines which components of the $\mathscr{A}^{\alpha\beta}$ and $\mathscr{D}^{\alpha\beta}$ tensors may be finite, and which ones will take the same value. For example, in cubic systems $\mathscr{A}^{\alpha\beta}$ and $\mathscr{D}^{\alpha\beta}$ are both constant matrices with the above definitions. For a list of Lifshitz invariants with finite $\mathscr{D}^{\alpha\beta}$ components in different symmetry classes see e.g. Refs.~\cite{Bogdanov,Li}.

The spin wave frequencies are calculated analogously to the atomistic model, by constructing the equation of motion\cite{Landau}
\begin{eqnarray}
\partial_{t}\boldsymbol{S}=\frac{\gamma}{\mathscr{M}}\boldsymbol{S}\times\frac{\delta F}{\delta \boldsymbol{S}},\label{eqnS32a}
\end{eqnarray}
then linearizing it in small deviations from the ferromagnetic state. Note that even if the micromagnetic model is used for the description of the system at room temperature, it is common practice to use Eq.~(\ref{eqnS32a}) for the calculation of the spectrum (see e.g. Ref.~\cite{Belmeguenai}), where the effect of temperature is only included in the interaction coefficients. A more accurate inclusion of finite-temperature effects in micromagnetic models is given by the Landau--Lifshitz--Bloch equation\cite{Chubykalo-Fesenko}. 

%To illustrate the correspondence between Eq.~(\ref{eqn2}) and Eq.~(\ref{eqn18}), note that both reproduce the Larmor frequency $\gamma B^{z}$ in the noninteracting case.

\section{Square lattice\label{secS3}}

For the model calculations we considered a ferromagnetic monolayer on the $(001)$ surface of a cubic lattice with $C_{4\textrm{v}}$ symmetry. 
% in the absence of external magnetic field. 
In this symmetry class, the micromagnetic exchange interaction corresponds to a constant tensor ($\mathscr{A}^{zz}=\mathscr{A}^{xx}$), while the Dzyaloshinsky--Moriya interaction tensor may be characterized by the single value $\mathscr{D}^{xz}=-\mathscr{D}^{zx}$.

The complete two-dimensional spin wave spectrum of the system with the magnetic field applied along the $z$ direction is given by
\begin{eqnarray}
\frac{\mu_{s}}{\gamma}\omega_{\boldsymbol{k}}\left(T\right)=&&2\mathcal{J}^{z}\left(1-\cos\left(k^{z}a\right)\right)+2\mathcal{J}^{x}\left(1-\cos\left(k^{x}a\right)\right)\nonumber
\\
&&+2\mathcal{D}\sin\left(k^{x}a\right)+2\mathcal{K}^{zz}+\mu_{s}B^{z},\label{eqnS33}
\end{eqnarray}
with
\begin{eqnarray}
\mathcal{J}^{z,x}&=&mJ+m^{2}J\:\textrm{Re}I^{z,x},\label{eqnS34}
\\
\mathcal{D}&=&mD+m^{2}J\:\textrm{Im}I^{x},\label{eqnS35}
\\
\mathcal{K}^{zz}&=&D\:\textrm{Im}I^{x}.\label{eqnS36}
\end{eqnarray}
Note that $\mathcal{J}$ in Eq.~(\ref{eqn22}) is denoted by $\mathcal{J}^{x}$ here for clarity.

The expressions $I^{z,x}$ appearing in Eqs.~(\ref{eqnS34})-(\ref{eqnS36}) account for the correlation corrections, and may be expressed for an infinite lattice by the formulae
\begin{eqnarray}
I^{z,x}=&&\frac{1}{2m}\left<S_{i+\delta_{z,x}}^{+}S_{i}^{-}\right>\nonumber
\\
=&&\left(\frac{a}{2\pi}\right)^{2}\int_{\textrm{BZ}}\textrm{e}^{-\textrm{i}k^{z,x}a}\frac{\gamma}{\mu_{s}}\frac{k_{\textrm{B}}T}{\omega_{\boldsymbol{k}}\left(T\right)}\textrm{d}^{2}\boldsymbol{k},\label{eqnS37}
\end{eqnarray}
where $i+\delta_{z,x}$ denotes the nearest neighbors of site $i$ along the positive $z$ and $x$ directions, respectively.

At zero temperature, from Eqs.~(\ref{eqnS34})-(\ref{eqnS36}) one obtains $\mathcal{J}^{z}=\mathcal{J}^{x}=J$, $\mathcal{D}=D$, and $\mathcal{K}^{zz}=0$, which satisfy the $C_{4\textrm{v}}$ symmetry of the system. However, it should be noted that at finite temperature one has $\mathcal{J}^{z}\neq\mathcal{J}^{x}$ and $\mathcal{K}^{zz}\neq 0$. This is caused by the simultaneous presence of the magnetic field along the $z$ direction which breaks the $C_{4\textrm{v}}$ symmetry and the Dzyaloshinsky--Moriya interaction.
% Since the temperature, the applied external magnetic field and the Dzyaloshinsky--Moriya interaction are all significantly weaker than the Heisenberg interactions in most systems, the unconventional deviations $\mathcal{J}^{z}-\mathcal{J}^{x}$ and $\mathcal{K}^{zz}$ are generally difficult to detect in the simulations or the experiments -- for further details see Appendix~\ref{secS4}.

Regarding Eq.~(\ref{eqnS37}), even at finite temperature $B$ and $D$ do not break the $\omega_{k^{z},k^{x}}=\omega_{-k^{z},k^{x}}$ symmetry of the spectrum, which implies that $I^{z}$ will remain real at all temperatures. In the main text, we have used the fact that the temperature dependence of the Heisenberg $\mathcal{J}=\mathcal{J}^{x}$ and the Dzyaloshinsky--Moriya $\mathcal{D}$ exchange interactions is exactly the same within Green's function formalism in the present system. This can be proven by introducing the simplified notations
\begin{eqnarray}
k'=&&k^{x}a,\label{eqnS38}
\\
a=&&2\mathcal{J}^{x}+2\mathcal{J}^{z}\left(1-\cos\left(k^{z}a\right)\right)+2\mathcal{K}^{zz}+\mu_{s}B^{z},\label{eqnS39}
\\
b=&&2\mathcal{J}^{x},\label{eqnS40}
\\
c=&&2\mathcal{D},\label{eqnS41}
\end{eqnarray}
and calculating the one-dimensional integrals
\begin{align}
\textrm{Re}I_{\textrm{1D}}^{x}=&\int_{0}^{2\pi}\frac{a\cos k'-b\cos^{2}k'}{a^{2}-c^{2}-2ab\cos k'+\left(b^{2}+c^{2}\right)\cos^{2}k'}\textrm{d}k',\label{eqnS43}
\\
\textrm{Im}I_{\textrm{1D}}^{x}=&\int_{0}^{2\pi}\frac{c-c\cos^{2}k'}{a^{2}-c^{2}-2ab\cos k'+\left(b^{2}+c^{2}\right)\cos^{2}k'}\textrm{d}k'.\label{eqnS44}
\end{align}

$I^{x}$ in Eq.~(\ref{eqnS37}) may be obtained from $I_{\textrm{1D}}^{x}$ by performing the integral over $k^{z}$ and multiplying by a constant factor.

By introducing
\begin{eqnarray}
\lambda_{\pm}=\frac{ab\pm\sqrt{a^{2}b^{2}-\left(a^{2}-c^{2}\right)\left(b^{2}+c^{2}\right)}}{b^{2}+c^{2}},\label{eqnS45}
\end{eqnarray}
Eqs.~(\ref{eqnS43})-(\ref{eqnS44}) may be expressed analytically as
\begin{eqnarray}
\textrm{Re}I_{\textrm{1D}}^{x}=&&-\frac{2\pi}{b^{2}+c^{2}}\Bigg(b+\frac{\lambda_{+}\left(a-b\lambda_{+}\right)}{\left(\lambda_{+}-\lambda_{-}\right)\sqrt{\lambda_{+}^{2}-1}}\nonumber
\\
&&-\frac{\lambda_{-}\left(a-b\lambda_{-}\right)}{\left(\lambda_{+}-\lambda_{-}\right)\sqrt{\lambda_{-}^{2}-1}}\Bigg),\label{eqnS46}
\\
\textrm{Im}I_{\textrm{1D}}^{x}=&&-\frac{2\pi}{b^{2}+c^{2}}\Bigg(c+\frac{c\left(1-\lambda^{2}_{+}\right)}{\left(\lambda_{+}-\lambda_{-}\right)\sqrt{\lambda_{+}^{2}-1}}\nonumber
\\
&&-\frac{c\left(1-\lambda^{2}_{-}\right)}{\left(\lambda_{+}-\lambda_{-}\right)\sqrt{\lambda_{-}^{2}-1}}\Bigg).\label{eqnS47}
\end{eqnarray}

It can be shown by algebraic transformations that
\begin{eqnarray}
\frac{\textrm{Im}I_{\textrm{1D}}^{x}}{\textrm{Re}I_{\textrm{1D}}^{x}}=\frac{\textrm{Im}I^{x}}{\textrm{Re}I^{x}}=\frac{\mathcal{D}}{\mathcal{J}^{x}}.\label{eqnS48}
\end{eqnarray}

Finally, substituting Eq.~(\ref{eqnS48}) into Eqs.~(\ref{eqnS34})-(\ref{eqnS35}) and calculating the ratio of the latter two yields
\begin{eqnarray}
\frac{\mathcal{D}}{\mathcal{J}^{x}}=\frac{D}{J},\label{eqnS49}
\end{eqnarray}
meaning that the ratio of the Dzyaloshinsky--Moriya and Heisenberg exchange interactions indeed does not depend on the temperature. We briefly mention that the exact equality only holds for only nearest-neighbor interactions in the model. Finally, substituting Eq.~(\ref{eqnS48}) into Eq.~(\ref{eqnS36}) yields Eq.~(\ref{eqn27a}) for the correspondence between the induced anisotropy and the Heisenberg exchange interaction.

\section{Monte Carlo simulations\label{secS4}}

We performed the Monte Carlo simulations on an $N=64\times64$ square lattice with periodic boundary conditions for the Hamiltonian illustrated in Fig.~\ref{fig1}. We used the single-spin Metropolis algorithm to update the spin directions, with sweeping over the whole lattice at every Monte Carlo step. From the simulations we extracted the thermal equilibrium quantities $m=\frac{1}{N}\sum_{i}\left<S_{i}^{z}\right>$ and $\left<S^{-}_{\boldsymbol{k}}S^{+}_{-\boldsymbol{k}}\right>$ from the lattice Fourier transform of the spin configurations. For the data displayed in Fig.~\ref{fig2}, we initialized the system in the ferromagnetic state, increased the temperature from $k_{\textrm{B}}T=0.05$ to $k_{\textrm{B}}T=1.50$ in steps of $k_{\textrm{B}}\Delta T=0.05$, performed thermalization for $2\times10^{5}$ Monte Carlo steps, and averaged the observables during the Monte Carlo evolution over $10^{5}$ configurations at $10^{3}$ Monte Carlo step distance from each other.
%, and finally averaged over $100$ independent realizations of the process.

After determining the observables, we calculated the spin wave frequencies from Eq.~(\ref{eqn27}), extracted the spectrum along the $x$ direction, and fitted the results with a function of the form Eq.~(\ref{eqn22}) to extract the temperature-dependent parameters. We compared the results to the prediction of Green's function theory, also calculated for an $N=64\times64$ lattice. Increasing the lattice size to $N=128\times128$ in the theoretical calculations modified the interaction parameters on the order of $10^{-10}$, well below the precision of the simulations and indicating that finite-size effects are negligible for the considered problem.

%It is expected that Green's function theory provides a satisfactory agreement with the simulations not only close to zero temperature, but over a wide range in magnetization. Figure~\ref{figS2} demonstrates that this is indeed the case. For these simulations we used the same procedure as in the case of Fig.~\ref{fig2}, apart from averaging over $100$ independent realizations. Note that on this scale the random phase approximation or mean-field model is also relatively close to the simulation results, since they also correctly describe the high-temperature behavior. The most pronounced deviations due to transversal fluctuations may be observed at lower temperature. 

%\begin{figure}
%\centering
%\includegraphics[width=\columnwidth]{K0D02B01longnew.eps}
%\caption{Effective temperature-dependent coefficients as a function of magnetization as in Fig.~\ref{fig2}, over a wider range in temperature ($0.0\le k_{\textrm{B}}T\le 1.0$) and magnetization.\label{figS2}}
%\end{figure}

We also calculated the spin wave spectrum along the $z$ direction from the simulations in order to confirm that it remains symmetric, and to extract the $\mathcal{J}^{z}$ parameter discussed in Appendix~\ref{secS3}. However, we were not successful in determining a deviation between $\mathcal{J}^{z}$ and $\mathcal{J}^{x}$ that would be significant compared to the uncertainty of the parameters obtained from the simulations by the fitting procedure.
% This indicates that the symmetry breaking created by the simultaneous presence of thermal excitations, external field and Dzyaloshinsky--Moriya interactions is indeed weak. Nevertheless, this symmetry breaking can be observed in the anisotropy term induced by the Dzyaloshinsky--Moriya interaction in Fig.~\ref{fig3}. This value is simpler to determine from the simulations, since it only requires the determination of the spin wave frequency at $\boldsymbol{k}=\boldsymbol{0}$ instead of a parameter fitting.
% As illustrated in Fig.~\ref{fig3}, it was possible to demonstrate the presence of the anisotropy term induced by the Dzyaloshinsky--Moriya interaction, because this only requires the determination of the spin wave frequency at $\boldsymbol{k}=\boldsymbol{0}$ instead of a parameter fitting. Although Green's function theory correctly predicts an increasing gap with temperature, it significantly overestimates its magnitude. Note that this increase with the temperature is opposite to the effect of considering an actual anisotropy term in the atomistic Hamiltonian, which would weaken with temperature.

%\begin{figure}
%\centering
%\includegraphics[width=\columnwidth]{K0D02B01gapnew.eps}
%\caption{Energy of the spin wave at wave vector $\boldsymbol{k}=\boldsymbol{0}$ as a function of temperature. The figure shows the comparison between the results of Monte Carlo simulations (MC), Green's function theory (GF) and the random phase approximation (RPA). The dimensionless interaction parameters are $J=1, D=-0.2, B^{z}=0.1, \mu_{s}=1$.\label{fig3}}
%\end{figure}


\begin{thebibliography}{1}

	\bibitem{Dzyaloshinsky} I. Dzyaloshinsky, J. Phys. Chem. Solids \textbf{4}, 241 (1958).

	\bibitem{Moriya} T. Moriya, Phys. Rev. Lett. \textbf{4}, 228 (1960).

	\bibitem{Thiaville} A. Thiaville, S. Rohart, E. Ju\'{e}, V. Cros, and A. Fert, Europhys. Lett. \textbf{100}, 57002 (2012).

	\bibitem{Ryu} K.-S. Ryu, L. Thomas, S.-H. Yang, and S. Parkin, Nat. Nanotechnol. \textbf{8}, 527 (2013).

	\bibitem{Bode} M. Bode, M. Heide, K. von Bergmann, P. Ferriani, S. Heinze, G. Bihlmayer, A. Kubetzka, O. Pietzsch, S. Bl\"{u}gel, and R. Wiesendanger, Nature (London) \textbf{447}, 190 (2007).

	\bibitem{Meckler} S. Meckler, N. Mikuszeit, A. Pressler, E. Y. Vedmedenko, O. Pietzsch, and R. Wiesendanger, Phys. Rev. Lett. \textbf{103}, 157201 (2009).

	\bibitem{Bogdanov} A. N. Bogdanov and D. A. Yablonski\u{i}, Zh. Eksp. Teor. Fiz. \textbf{95}, 178 (1989) [Sov. Phys. JETP \textbf{68}, 101 (1989)].

	\bibitem{Muhlbauer} S. M\"{u}hlbauer, B. Binz, F. Jonietz, C. Pfleiderer, A. Rosch, A. Neubauer, R. Georgii, and P. B\"{o}ni, Science \textbf{323}, 915 (2009).

	\bibitem{Yu2} X. Z. Yu, N. Kanazawa, Y. Onose, K. Kimoto, W. Z. Zhang, S. Ishiwata, Y. Matsui, and Y. Tokura, Nat. Mater. \textbf{10}, 106 (2010).

	\bibitem{Allwood} D. A. Allwood, G. Xiong, C. C. Faulkner, D. Atkinson, D. Petit, and R. P. Cowburn, Science \textbf{309}, 1688 (2005).

	\bibitem{Parkin} S. S. P. Parkin, M. Hayashi, and L. Thomas, Science \textbf{320}, 190 (2008).

	\bibitem{Fert} A. Fert, V. Cros, and J. Sampaio, Nat. Nanotechnol. \textbf{8}, 152 (2013).

	\bibitem{Melcher} R. L. Melcher, Phys. Rev. Lett. \textbf{30}, 125 (1973).

	\bibitem{Udvardi} L. Udvardi and L. Szunyogh, Phys. Rev. Lett. \textbf{102}, 207204 (2009).

	\bibitem{Coldea} R. Coldea, D. A. Tennant, K. Habicht, P. Smeibidl, C.~Wolters, and Z. Tylczynski, Phys. Rev. Lett. \textbf{88}, 137203 (2002).

	\bibitem{Sato} T. J. Sato, D. Okuyama, T. Hong, A. Kikkawa, Y.~Taguchi, T.-h. Arima, and Y. Tokura, Phys. Rev. B \textbf{94}, 144420 (2016).

	\bibitem{Nembach} H. T. Nembach, J. M. Shaw, M. Weiler, E. Ju\'{e}, and Th.~J.~Silva, Nat. Phys. \textbf{11}, 825 (2015).

	\bibitem{Belmeguenai} M. Belmeguenai, J.-P. Adam, Y. Roussign\'{e}, S. Eimer, T. Devolder, J.-V. Kim, S. M. Cherif, A. Stashkevich, and A. Thiaville, Phys. Rev. B \textbf{91}, 180405 (2015).

	\bibitem{Zakeri} Kh. Zakeri, Y. Zhang, J. Prokop, T.-H. Chuang, N. Sakr, W. X. Tang, and J. Kirschner, Phys. Rev. Lett. \textbf{104}, 137203 (2010).

	\bibitem{Zakeri3} Kh. Zakeri, T.-H. Chuang, A. Ernst, L. M. Sandratskii, P. Buczek, H. J. Qin, Y. Zhang, and J. Kirschner, Nat. Nanotechnol. \textbf{8}, 853 (2013).

	\bibitem{Zakeri2} Kh. Zakeri, J. Prokop, Y. Zhang, and J. Kirschner, Surf. Sci. \textbf{630}, 311 (2014).

	\bibitem{Lee} J. M. Lee, C. Jang, B.-C. Min, S.-W. Lee, K.-J. Lee, and J. Chang, Nano Lett. \textbf{16}, 62 (2016).

	\bibitem{Garcia-Sanchez} F. Garcia-Sanchez, P. Borys, R. Soucaille, J.-P. Adam, R. L. Stamps, and J.-V. Kim, Phys. Rev. Lett. \textbf{114}, 247206 (2015).

	\bibitem{WYu} W. Yu, J. Lan, R. Wu, and J. Xiao, Phys. Rev. B \textbf{94}, 140410(R) (2016).

	\bibitem{Fert2} A. Fert and P. M. Levy, Phys. Rev. Lett. \textbf{44}, 1538 (1980).

	\bibitem{Levy} P. M. Levy and A. Fert, Phys. Rev. B \textbf{23}, 4667 (1981).

	\bibitem{Cubukcu} M. Cubukcu, J. Sampaio, K. Bouzehouane, D. Apalkov, A. V. Khvalkovskiy, V. Cros, and N. Reyren, Phys. Rev. B \textbf{93}, 020401(R) (2016).

	\bibitem{Hagemeister} J. Hagemeister, N. Romming, K. von Bergmann, E. Y. Vedmedenko, and R. Wiesendanger, Nat. Commun. \textbf{6}, 8455 (2015).

	\bibitem{Oike} H. Oike, A. Kikkawa, N. Kanazawa, Y. Taguchi, M. Kawasaki, Y. Tokura, and F. Kagawa, Nat. Phys. \textbf{12}, 62 (2016).

	\bibitem{Rozsa2} L. R\'{o}zsa, E. Simon, K. Palot\'{a}s, L. Udvardi, and L. Szunyogh, Phys. Rev. B \textbf{93}, 024417 (2016).

	\bibitem{Chubykalo-Fesenko} O. Chubykalo-Fesenko, U. Nowak, R. W. Chantrell, and D. Garanin, Phys. Rev. B \textbf{74}, 094436 (2006).

	\bibitem{Atxitia2} U. Atxitia, D. Hinzke, and U. Nowak, J. Phys. D: Appl. Phys. \textbf{50}, 033003 (2016).

	\bibitem{Schlickeiser} F. Schlickeiser, U. Ritzmann, D. Hinzke, and U. Nowak, Phys. Rev. Lett. \textbf{113}, 097201 (2014).

	\bibitem{Selzer} S. Selzer, U. Atxitia, U. Ritzmann, D. Hinzke, and U. Nowak, Phys. Rev. Lett. \textbf{117}, 107201 (2016).

	\bibitem{Eggebrecht} T. Eggebrecht, M. M\"{o}ller, J. G. Gatzmann, N. Rubiano da Silva, A. Feist, U. Martens, H. Ulrichs, M. M\"{u}nzenberg, C. Ropers, and S. Sch\"{a}fer, Phys. Rev. Lett. \textbf{118}, 097203 (2017).

	\bibitem{Finazzi} M. Finazzi, M. Savoini, A. R. Khorsand, A. Tsukamoto, A. Itoh, L. Duo, A. Kirilyuk, T. Rasing, and M. Ezawa, Phys. Rev. Lett. \textbf{110}, 177205 (2013).

	\bibitem{Callen2} H. B. Callen and E. Callen, J. Phys. Chem. Sol. \textbf{27}, 1271 (1966).

	\bibitem{Okamoto} S. Okamoto, N. Kikuchi, O. Kitakami, T. Miyazaki, Y. Shimada, and K. Fukamichi, Phys. Rev. B \textbf{66}, 024413 (2002).

	\bibitem{Asselin} P. Asselin, R. F. L. Evans, J. Barker, R. W. Chantrell, R. Yanes, O. Chubykalo-Fesenko, D. Hinzke, and U. Nowak, Phys. Rev. B \textbf{82}, 054415 (2010).

	\bibitem{Atxitia} U. Atxitia, D. Hinzke, O. Chubykalo-Fesenko, U. Nowak, H. Kachkachi, O. N. Mryasov, R. F. Evans, and R.~W.~Chantrell, Phys. Rev. B \textbf{82}, 134440 (2010).

	\bibitem{Heider} F. Heider and W. Williams, Geophys. Res. Lett. \textbf{15}, 184 (1988).

	\bibitem{Barker} J. Barker and O. A. Tretiakov, Phys. Rev. Lett. \textbf{116}, 147203 (2016).

	\bibitem{Kim} S. Kim, K. Ueda, G. Go, P.-H. Jang, K.-J. Lee, A. Belabbes, A. Manchon, M. Suzuki, Y. Kotani, T. Nakamura, K. Nakamura, T. Koyama, D. Chiba, K. Yamada, D.-H. Kim, T. Moriyama, K.-J. Kim, and T. Ono, arXiv:1704.02900 (2017).

	\bibitem{Hasselberg} G. Hasselberg, R. Yanes, D. Hinzke, P. Sessi, M. Bode, L. Szunyogh, and U. Nowak, Phys. Rev. B \textbf{91}, 064402 (2015).

	\bibitem{Sonntag} A. Sonntag, J. Hermenau, S. Krause, and R. Wiesendanger, Phys. Rev. Lett. \textbf{113}, 077202 (2014).

	\bibitem{Sessi} P. Sessi, N. P. Guisinger, J. R. Guest, and M. Bode, Phys. Rev. Lett. \textbf{103}, 167201 (2009).

	\bibitem{Bergmann} K. von Bergmann, M. Bode, and R. Wiesendanger, J.~Magn.~Magn.~Mater. \textbf{305}, 279 (2006).

	\bibitem{Shibata} K. Shibata, A. Kov\'{a}cs, N. S. Kiselev, N. Kanazawa, R.~E.~Dunin-Borkowski, and Y. Tokura, Phys. Rev. Lett. \textbf{118}, 087202 (2017).

	\bibitem{Rozsa} L. R\'{o}zsa, L. Udvardi, and L. Szunyogh, J. Phys.: Condens. Matter \textbf{25}, 506002 (2013).

	\bibitem{Callen} H. B. Callen, Phys. Rev. \textbf{130}, 890 (1963).

	\bibitem{Bastardis} R. Bastardis, U. Atxitia, O. Chubykalo-Fesenko, and H.~Kachkachi, Phys. Rev. B \textbf{86}, 094415 (2012).

	\bibitem{Campana} L. S. Campana, A. Cavallo, L. De Cesare, U. Esposito, and A. Naddeo, Physica A \textbf{391}, 1087 (2012).

	\bibitem{Nakamura} T. Nakamura and M. Bloch, Phys. Rev. \textbf{132}, 2528 (1963).

	\bibitem{You} Ch.-Y. You, J. Appl. Phys. \textbf{116}, 053902 (2014).

	\bibitem{Cho} J. Cho, N.-H. Kim, S. Lee, J.-S. Kim, R. Lavrijsen, A. Solignac, Y. Yin, D.-S. Han, N. J. J. van Hoof, H. J. M. Swagten, B. Koopmans, and Ch-Y. You, Nat. Commun. \textbf{6}, 7635 (2015).

	\bibitem{Tyablikov} S. V. Tyablikov, Ukr. Mat. Zh. \textbf{11}, 287 (1959).

	\bibitem{Anderson} P. W. Anderson, Phys. Rev. \textbf{115}, 2 (1959).

	\bibitem{Szilva} A. Szilva, M. Costa, A. Bergman, L. Szunyogh, L. Nordstr\"{o}m, and O. Eriksson, Phys. Rev. Lett. \textbf{111}, 127204 (2013).

	\bibitem{Swendsen} R. H. Swendsen, Phys. Rev. B \textbf{5}, 116 (1972).

	\bibitem{Mermin} N. D. Mermin and H. Wagner, Phys. Rev. Lett. \textbf{17}, 1133 (1966).

	\bibitem{Vida} Gy. J. Vida, E. Simon, L. R\'{o}zsa, K. Palot\'{a}s, and L. Szunyogh, Phys. Rev. B \textbf{94}, 214422 (2016).

	%

	\bibitem{Frobrich} P. Fr\"{o}brich and P. J. Kuntz, Phys. Rep. \textbf{432}, 223 (2006).

	\bibitem{Callen3} H. B. Callen and S. Shtrikman, Sol. State Commun. \textbf{3}, 5 (1965).

	\bibitem{Schweflinghaus} B. Schweflinghaus, B. Zimmermann, M. Heide, G. Bihlmayer, and S. Bl\"{u}gel, Phys. Rev. B \textbf{94}, 024403 (2016).

	\bibitem{Li} W. Li, Ch. Jin, R. Che, W. Wei, L. Lin, L. Zhang, H. Du, M. Tian, and J. Zang, Phys. Rev. B \textbf{93}, 060409(R) (2016).

	\bibitem{Landau} L. D. Landau, E. M. Lifshitz, \textit{Theoretical Physics, Vol. IX, Statistical Physics Pt. 2} (Nauka, Moscow, 1978).

%	\bibitem{Brown} W. F. Brown, \textit{Micromagnetics} (Wiley, New York, 1963).

\end{thebibliography}
\end{document}